\documentclass[10pt, aps,prd, nofootinbib, preprintnumbers, reprint, 
  twocolumn, superscriptaddress]{revtex4-1}

\usepackage[pdftex]{graphicx}
\usepackage{amssymb, amsmath, bm}
\usepackage{microtype}
\usepackage{etoolbox}
\usepackage{mathtools}
\apptocmd{\sloppy}{\hbadness 10000\relax}{}{}

\usepackage{xcolor}

\usepackage[colorlinks,linktocpage,hypertexnames=false]{hyperref}
\hypersetup{colorlinks=true, citecolor=blue}

\def\e{{\rm e}}

\def\l{\left(}
\def\r{\right)}

\newcommand{\be}{\begin{equation}}
\newcommand{\ee}{\end{equation}}
\newcommand{\ba}{\begin{align}}
\newcommand{\ea}{\end{align}}
\newcommand{\bg}{\begin{gather}}
\newcommand{\eg}{\end{gather}}
\newcommand{\bseq}{\begin{subequations}}
\newcommand{\eseq}{\end{subequations}}

\makeatletter
\def\gsim{\compoundrel>\over\sim}
\def\lsim{\compoundrel<\over\sim}
\def\compoundrel#1\over#2{\mathpalette\compoundreL{{#1}\over{#2}}}
\def\compoundreL#1#2{\compoundREL#1#2}
\def\compoundREL#1#2\over#3{\mathrel
         {\vcenter{\hbox{$\m@th\buildrel{#1#2}\over{#1#3}$}}}}
\makeatother

\begin{document}

\preprint{INR-TH-2025-023}
\title{
Only slightly ``Amplifying nonresonant production of dark sector
  particles in scattering dominance regime''
} 

\author{S.V. Demidov}\email{demidov@ms2.inr.ac.ru}
\affiliation{Institute for Nuclear Research of the Russian Academy of
  Sciences, Moscow 117312, Russia}
\affiliation{Moscow Institute of Physics and Technology, Dolgoprudny
  141700, Russia}
\affiliation{Faculty of Physics, Moscow State University, Moscow 119991, Russia}
\author{D.S. Gorbunov}\email{gorby@ms2.inr.ac.ru}
\affiliation{Institute for Nuclear Research of the Russian Academy
  of Sciences, Moscow 117312, Russia}
\affiliation{Moscow Institute of Physics and Technology, Dolgoprudny
  141700, Russia}
\author{A.L. Polonski}\email{polonski@inr.ru}
\affiliation{Institute for Nuclear Research of the Russian Academy
  of Sciences, Moscow 117312, Russia}

\begin{abstract}
  In Ref.\,\cite{Du:2023zlt} it has been argued that production of
  dark photons---hypothetical massive vectors---in nuclear reactors via mixing with the visible photons is
considerably enhanced (by a factor of 10) due to Compton scattering of
the latter. We revisit the production of dark photons in the reactor
environment and find that although the scattering of photons indeed leads
to a larger number of produced dark photons, the overall enhancement is
considerably smaller than what has been obtained in~\cite{Du:2023zlt}. Our
findings are validated using GEANT simulations, which take into
account oscillations between ordinary and visible photons and
interaction of the latter with matter. The correction to the limit on mixing parameter between dark and visible photons is expected to be below 30\% for all masses. It's application must be accompanied with an update on the original spectrum of photons produced in nuclear reactions inside the nuclear reactor.    
\end{abstract}

\maketitle


{\bf 1.} Massive vector boson interacting with the Standard Model (SM)
particles via mixing with photon (so called vector portal) is predicted in
many extensions of SM, providing with an interesting phenomenology
being either dark matter candidate or mediator to a hidden sector, see 
e.g.~\cite{Fabbrichesi:2020wbt} for a review. 
The relevant part of the model largangian has the
form~\cite{Holdom:1985ag} 
\be
\label{eq1}
{\cal L} = - \frac{1}{4}\, F_{\mu\nu}^2 - \frac{1}{4}\, X_{\mu\nu}^2
        - \frac{\epsilon}{2}X_{\mu\nu}F^{\mu\nu} + \frac{m_X^2}{2}
        X_\mu^2 - e A_\mu j^\mu_{em}\,,
\ee
where $F_{\mu\nu} = \partial_\mu A_\nu - \partial_\nu A_\mu$ and
$X_{\mu\nu} = \partial_\mu X_\nu - \partial_\nu X_\mu$  with
$A_\mu$ and $X_\nu$ being the vector fields of  visible ($\gamma$) and
dark ($X$) photons, respectively. The former field interacts with the
electromagnetic current $j_\mu^{em}$; $m_{X}$ and $\epsilon\ll1 $ are
 model parameters. 

Our subject is the production of a dark photon via its mixing with the
ordinary photon (see Eq.~\eqref{eq1}) in the medium. Specifically, we
consider the hidden photon lighter than about 1~MeV. For this
mass range the strongest of all bounds are the astrophysical ones, 
see e.g.~\cite{Caputo:2021eaa}. Still, direct searches play an
important role typically presenting more robust limits.
In this respect searches 
for dark photons, emerged from a nuclear reactor 
~\cite{Park:2017prx,Danilov:2018bks,Demidov:2018odn}, present an  
interesting avenue for these studies. Photons of energies
$E\sim1\!-\!10$\,MeV, produced abundantly in the reactor core, can
oscillate to dark photons if kinematically allowed. The latter then
propagate freely through the reactor and its shielding to give
a signal in a distant
detector. 
The energy spectrum of ordinary photons produced in the reactor is rather complicated due to their numerous sources. In what follows we use for
the high energy tail, $E\gtrsim 0.2$\,MeV, of the photon  
spectrum the estimate presented in Ref.~\cite{FRJ} and  based on the most relevant nuclear reactions accompanying with $\gamma$-production. It has the
exponentially decaying form  
\be
\label{eq:reactor}
\frac{dN_\gamma}{dE}=0.58\times 10^{21}\times \frac{T}{\text{\rm GW}}
\times \e^{-\frac{E}{0.91\,\text{\rm MeV}}}\,
\ee
in units of photons/(s$\times$MeV), with $T$ being the reactor thermal power. 
Let us note that other directions of direct searches for a light dark
photon include hunts for light dark photons at colliders, in
particular, in future experiments~\cite{Seo:2020dtx,NEON:2024bpw}. 

{\bf 2.} Oscillations between visible and hidden photons can be
described~\cite{Redondo:2015iea,Redondo:2008aa,Redondo:2013lna} by the 
following Hamiltonian 
\begin{equation}
\label{eq2}
H = \frac{1}{2E}
\left(\begin{array}{cc}
  \epsilon^2 m_X^2 + m_\gamma^2 & -\epsilon m_X^2 \\
  -\epsilon m_X^2 & m_X^2
\end{array}\right)\;.
\end{equation}
Here $m_\gamma$ is an effective mass of photon, which the latter
acquires due to its coherent forward scattering off free  electrons in 
the media~\cite{Braaten:1993jw}, $m_\gamma^2=4\pi\alpha
n_e/m_e$. The effective Hamiltonian \eqref{eq2} is justified in the
ultrarelativistic case, $E_\gamma\gg m_X,\,m_\gamma$.
The transition probability over the distance $L$ has the
form of standard oscillations
\be
\label{eq:osc}
P(\gamma\to X) = \sin^2{2\theta}\sin^2\left(\frac{\Delta
  m^2L}{4E}\right)\,, 
\ee
where
\be
\label{eq:tan}
\tan{2\theta} = \frac{2\epsilon m_X^2}{\Delta m^2}
\ee
with $\Delta m^2 \equiv m_X^2-m_\gamma^2$. Corresponding oscillation 
length scales with the parameters as $  L_{osc}\approx
2.5\,\text{\rm   cm} 
\times\frac{E}{1\,\text{\rm MeV}}\frac{\l 10\,\text{\rm
    eV}\r^2}{\Delta m^2}\,$.  

\begin{figure}[!htb]
    \includegraphics[width=9.0cm]{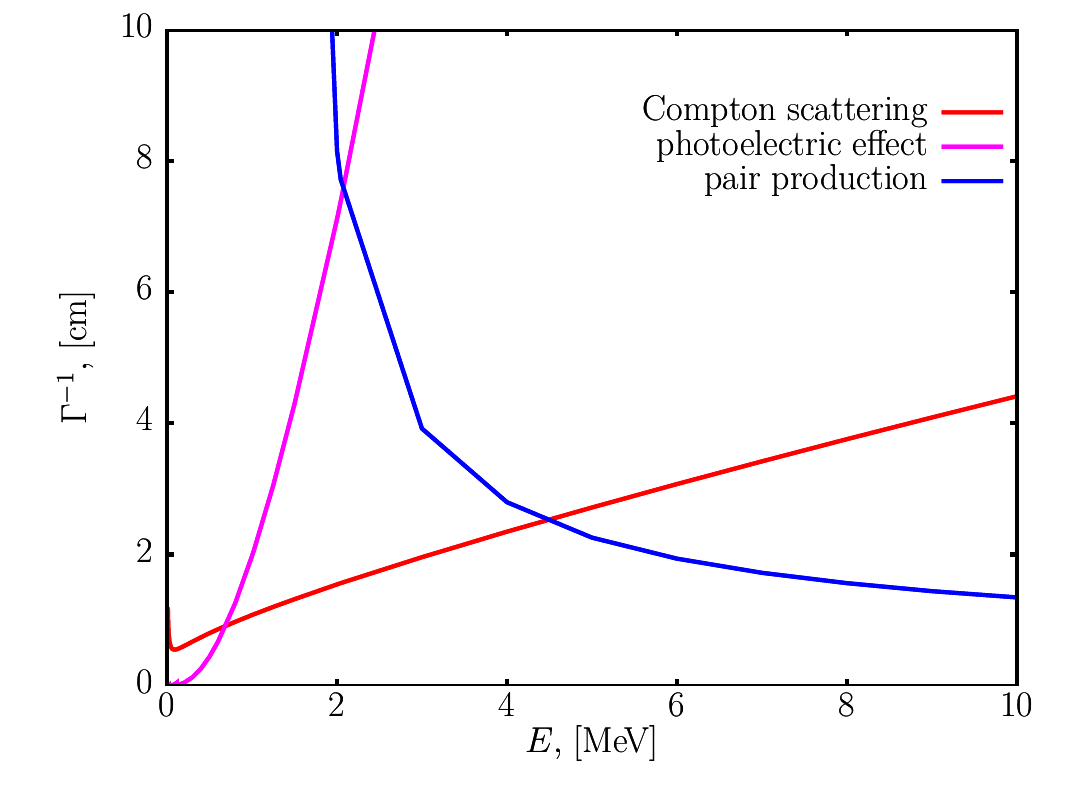}
    \caption{\label{fig:photon_proc}
Photon interaction lengths for Compton scattering,
atomic photoelectric effect and pair production processes in uranium
matter. Data are taken from~\cite{148746}.}   
\end{figure}

This simple oscillation picture becomes more complicated when
incoherent interactions of photon with the media are taken into
account\footnote{We stress that Eq.~\eqref{eq:reactor} corresponds to
energy spectrum of the {\it primary} photons, i.e. photons emitted in
nuclear reactions and radioactive decays in the reactor.}. For the  
relevant photon energy range the most important processes
include~\cite{ParticleDataGroup:2024cfk} a photoelectric absorption, 
a pair production and Compton scattering. 
The first two of them lead to photon disappearance (i.e. absorption by
the media), while the Compton scattering results in the production of a 
secondary photon, which propagates with a lower energy. 
Figure\,\ref{fig:photon_proc}
shows the absorption lengths for each of these processes for
uranium\footnote{Ref.~\cite{Du:2023zlt} takes thorium as a dominant
element of the reactor active zone. But the ratio
$\Gamma_{sca}/\Gamma_{abs}$ for uranium and thorium are almost the same.}
$U^{238}$ media calculated using Ref.\,\cite{148746}. As we see, the Compton
scattering is important in the photon energy range
$0.7-4.5$~MeV, where $\Gamma_{sca}/\Gamma_{abs}$ (with $\Gamma_{sca}$ and
$\Gamma_{abs}$ being inverse scattering and absorption lengths,
respectively) can reach a large factor of few (up to about
5). However, outside this energy range photon absorption processes
noticeably dominate. 

\begin{figure}[!htb]
    \includegraphics[width=9.0cm]{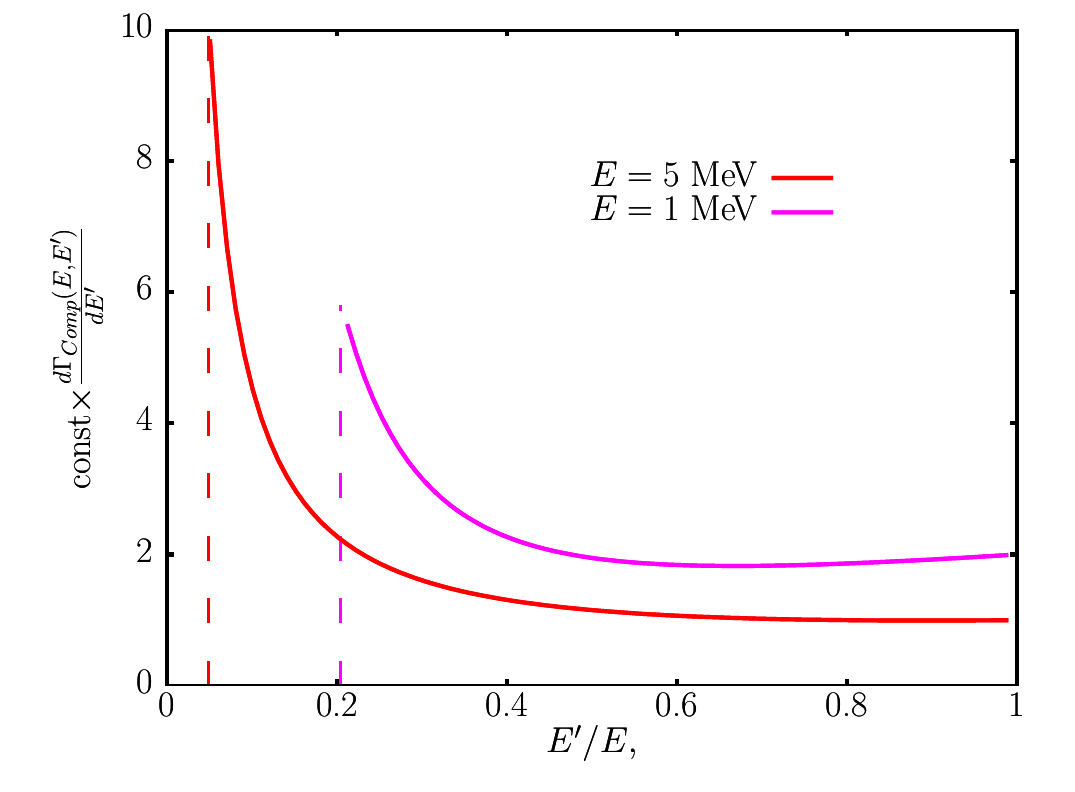}
    \caption{\label{fig:sec_photon}
Energy spectrum of outgoing photon after Compton scattering for
a set of initial photons energies.}   
\end{figure}

The energy spectrum of Compton-scattered photons has
the following form (see e.g.~\cite{LL4})
\be
\label{eq3}
\begin{split}
\frac{d\Gamma_{Comp}}{dE^\prime}(E,E^\prime) & \propto \frac{E}{E^\prime} + \frac{E^{\prime}}{E} +
\left(\frac{m_e}{E^\prime} - \frac{m_e}{E}\right)^2 \\ &  
-2\left(\frac{m_e}{E^\prime} - \frac{m_e}{E}\right),
\end{split}
\ee
where $E$ and $E^\prime$ are the initial and final photon energies,
respectively. The energy range of scattered photon reads
\be
\label{eq4}
\frac{E}{1+2E/m_e}\le E^\prime \le E\,.
\ee
Analyzing Eqs.~\eqref{eq3} and~\eqref{eq4} we see that the secondary
photons produced by the primary ones having relatively large
energies, i.e. $E\gg m_e$, exhibit energy spectrum strongly peaked at
lower part of 
the energy range~\eqref{eq4}. Increasing energy of the primary photon
results in further flattening of the flat part and further raising of the peak part of the photon spectrum in the reactor, see Fig.~\ref{fig:sec_photon}
for illustration. 

{\bf 3.} Early calculations~\cite{Danilov:2018bks} of dark photon production in
a reactor took into account photon absorption processes only. In this
case the produced dark photon has the same energy as the parent visible
photon and conversion probability has the form~\cite{Redondo:2015iea}
\be
\label{prob:media}
  P=\epsilon^2\times\frac{m_X^4}{\l\Delta m^2\r^2+ E^2 \Gamma_{abs}^2}\,  
\ee
for distances much larger than $\Gamma_{abs}^{-1}$. 
The authors of~\cite{Du:2023zlt} estimated the impact of
secondary photons on the dark photon production.
Their line of reasoning is as follows. Photon
scattering represents an act of measurement in which the pure state 
$a|\gamma, E\rangle + b|X, E\rangle$ collapses into the mixed
state described by the density matrix $\hat{\rho} =
|a|^2|\gamma, E^\prime\rangle\langle\gamma, E^\prime| +
|b|^2|X, E\rangle\langle X, E|$ with $|b|^2\sim
{\cal O}(\epsilon^2)$. The secondary photon having lower energy
$E^\prime$ has a chance to scatter again, resulting in the production of a 
dark photon with energy $E^\prime$. This evolution proceeds
until photon absorption or until $E^\prime < m_X$. Therefore, the
total number of produced dark photons is enhanced by a factor of order
of number of photon scatterings before absorption, which is of order
$\Gamma_{sca}/\Gamma_{abs}$.

However, as we see from Fig.~\ref{fig:photon_proc}, the total number of
photon scatterings can hardly be quantified by the 
$\Gamma_{sca}/\Gamma_{abs}$ because the latter has very strong energy
dependence. Figure\,\ref{fig:photon_proc} clearly demonstrates that
the Compton  scattering in a reactor may indeed have a noticeable
effect for the  photons having $E\lsim 4.5$~MeV, while for larger
energies one does not expect any perceptible contribution to dark
photon production from the 
secondary photons. This conclusion appears in a sharp contrast with the
results of Ref.~\cite{Du:2023zlt} where an enhancement of order 2 due
to the photon scattering was observed already for $E\approx 5$~MeV,
see Fig.~3 in Ref.~\cite{Du:2023zlt}. 
Next, the energy spectra of
Compton-scattered photons shown in Eqs.~\eqref{eq3},\eqref{eq4} and
Fig.~\ref{fig:sec_photon} hint that average photons with $E\gsim 
1$~MeV lose considerable amount of energy in a single scattering
event and therefore quickly degrade to very small energies. 
Furthermore, given the exponentially decaying with growing energy spectrum of
reactor photons (see Eq.~\eqref{eq:reactor}), a tenfold increase in
the dark photon spectrum from Compton scattering observed
in~\cite{Du:2023zlt} is highly improbable, as the number of
high-energy photons available for scattering is itself exponentially
smaller.  

The resolution of the problem relies on the observation made long
ago~\cite{Stodolsky:1986dx,Thomson:1991xq}, that the picture where 
interaction acts as a measurement described above (which can be
extended not only to the scattering but also to the absorption
processes) is not entirely accurate. As it is well known it indeed
leads to a correct description if the rate of  the measurement is
taken to be {\it one half} of the interaction  frequency of the
visible photon. Such a description is widely used, for instance, for
calculations of sterile neutrino production in a hot plasma (see
e.g.~\cite{Foot:1996qc}). Authors of Ref.~\cite{Du:2023zlt} considered
absorption and scattering processes on different footing: while
their expression for absorption probability is in agreement with the
above description, their consideration of scattering process follows
the incorrect interpretation of scattering as a measurement which
completely collapses the wave function, so the factor of one half is lost. 

{\bf 4.} Having the arguments presented above we revisit the dark
photon production in the reactor media by performing an accurate
calculation. We consider the density matrix of the visible-hidden
photon system
\be
\rho(E) = \left(\begin{array}{cc}
  \rho_{11}(E) & \rho_{12}(E)\\
  \rho_{12}^*(E) & \rho_{22}(E)
  \end{array}\right)\,,
\ee
which obeys the following equation~\cite{Sigl:1993ctk}
\be
\label{eq:density_matrix}
\frac{d\rho}{dr} = -i[H,\rho] - \frac{1}{2}\{G_{abs},\rho\} +
\frac{1}{2}G_{prod}\,.
\ee
Here we neglect irrelevant Bose enhancement and introduce $G_{abs} =
{\rm   diag}(\Gamma_{tot},0)$ with
$\Gamma_{tot}\equiv\Gamma_{abs}+\Gamma_{sca}$, $G_{prod} = {\rm
  diag}(Q_{prod},0)$ where  
\be
Q_{prod}(E) = \int_{E}^{\infty} dE^\prime
\frac{d\Gamma_{Comp}}{dE}(E^\prime,E)\rho_{11}(E^\prime)
\ee
describes the contribution of secondary Compton-scattered photons. 
According to Eq.~\eqref{eq:density_matrix}, the evolution equations for 
components of the density matrix read
\begin{align}
  \label{eq:system1}
& \frac{d\rho_{11}}{dr} & = &\;\;\;\; i\frac{\epsilon m_X^2}{2E}(\rho_{12}^* -
\rho_{12}) - \Gamma_{tot}\rho_{11} + \Gamma_{prod}\,,\\\nonumber
& \frac{d\rho_{12}}{dr} & = &\;\;\;  - i\frac{m_\gamma^2-(1-\epsilon^2)m_X^2}{2E}\rho_{12} \\ \label{eq:system2} & & & \;\;\;
+ i\frac{\epsilon m_X^2}{2E}(\rho_{22}-\rho_{11})
-\frac{1}{2}\Gamma_{tot}\rho_{11}\,,\\ \label{eq:system3}
& \frac{d\rho_{22}}{dr} & = &\;\;\;\; i\frac{\epsilon m_X^2}{2E}(\rho_{12}-\rho_{12}^*)\,.
\end{align}
If one solves this system with the initial conditions
\be
\rho_{11}\big|_{r=0} = \frac{dN_\gamma}{dE},\;\;
\rho_{12}\big|_{r=0}=\rho_{22}\big|_{r=0}=0\,,
\ee
then, assuming the size of the reactor active zone being much larger than the photon
absorption length, the energy spectra of produced dark photons
is given by
\be
\frac{dN_X}{dE} = \rho_{22}\big|_{r\to\infty}\,.
\ee
Without contribution from the secondary photons (i.e. for
$\Gamma_{prod}=0$) the
system~\eqref{eq:system1}--\eqref{eq:system3} can be easily solved
analytically order by order in $\epsilon$ resulting in
\be
\label{eq:dark}
\frac{dN_X}{dE} = \epsilon^2\times\frac{m_X^4}{\l\Delta m^2\r^2+ E^2
  \Gamma_{tot}^2} \times \frac{dN_\gamma}{dE}
\ee
up to corrections of the next order in $\epsilon$, which agrees with
the use of Eq.\eqref{prob:media} after a replacement
$\Gamma_{abs}\to\Gamma_{tot}$. Alternatively, the
spectrum~\eqref{eq:dark} can be reproduced via Monte-Carlo (MC)
simulation of photon propagation in which the first interaction of
photon is considered as a measurement resulting in a production of
dark photon with probability which is {\it one half} of the
probability~\eqref{eq:osc} while the scattered photons are neglected. 

\begin{figure}[!htb]
    \includegraphics[width=9.0cm]{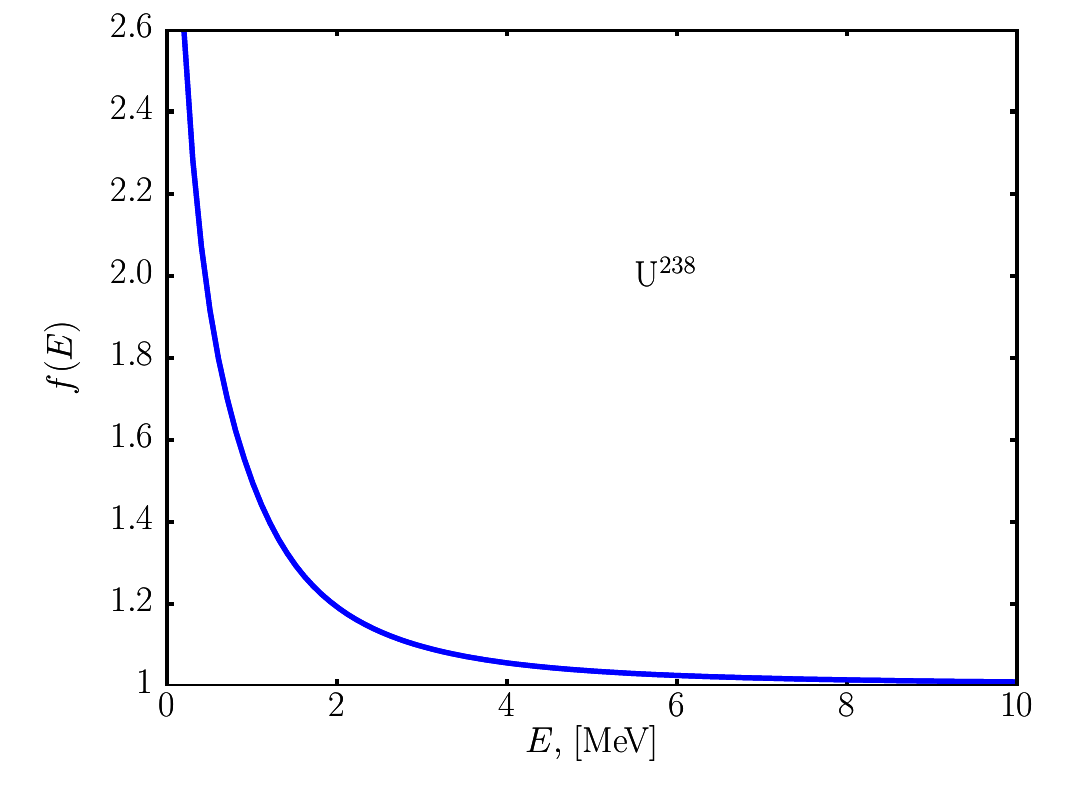}
    \caption{\label{fig:enhancement}
      Enhancement factor $f(E)$ for the dark photon production
      due to secondary photons from Compton scattering in media
      made of U$^{238}$.}   
\end{figure}

The secondary photons can be taken into account by either solving
numerically Eqs.~\eqref{eq:system1}--\eqref{eq:system3} with 
$\Gamma_{prod}$ or performing MC simulation with
scattered photons and with the factor {\it one half} in the
oscillation probability. Below we follow the second approach.  Direct
numerical integration of the
system~\eqref{eq:system1}--\eqref{eq:system3} is feasible only when
$L_{osc}$ is larger than or comparable to $\Gamma_{tot}^{-1}$, which
happens when $m_X$ is close to $m_\gamma$, where we use it to 
check the MC simulation results.
In the MC simulations we consider uranium $U^{238}$ media and propagate
photons taking into account the Compton scattering, atomic
photoelectric effect as well as $e^+$-$e^-$-pair production. 
We define the enhancement factor $f(E)$ as the ratio of produced dark
photon differential fluxes with and without secondary photons from
Compton scattering.
This enhancement factor is determined by the interactions of visible
photon with matter and is independent of the mass of the latter.
Figure~\ref{fig:enhancement}
shows $f(E)$ obtained from the MC simulations and, independently, by direct numerical solution of Eqs.~\eqref{eq:system1}--\eqref{eq:system3}. We see that the effect
of Compton-scattered photons is prominent at low energy part of 
the spectra only, in agreement with the qualitative
arguments presented above. Namely, the amplification is about 20\% at $E\simeq 2$\,MeV and reaches about 100\% at $E\sim 0.5$\,MeV. Since the dark photon flux is proportional to the quartic power of mixing, the limits obtained in Ref.\,\cite{Danilov:2018bks} can be improved by 30\% at best and only for sufficiently light dark photons. 

\begin{figure}
    \includegraphics[width=9.0cm]{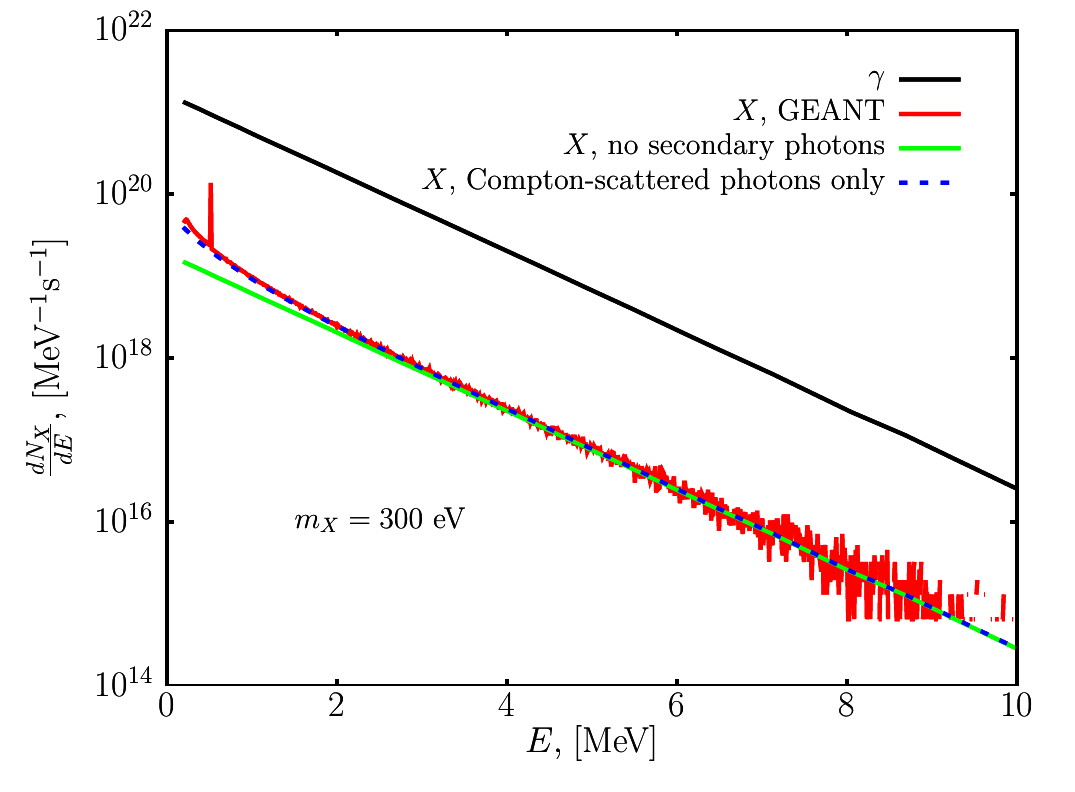}
    \caption{\label{fig:dp_GEANT}
The spectra of visible photons as well as dark photons calculated
without scattering effect, with Copmton-scattered photons and using
GEANT simulation. We assume $m_X=300$~eV, 
$\epsilon=0.01$ and 2.9~GW thermal power of the reactor. }   
\end{figure}

One may wonder about possible contribution to the dark photon
production from the secondary $e^+e^-$ pairs. To explore this effect
we turn to full Monte-Carlo simulation of the dark photon production in the reactor
media with GEANT (version 3.21). 
We introduced an additional process of dark photon production to the
simulation of electromagnetic processes in the GEANT code.
 In the simulations we adopt an U$^{238}$ active zone
of the reactor to be of a cubic geometry with a side 
length of 3~m surrounded by a large water shielding area.
The production of hidden photons has been introduced as an additional 
process to the simulation of electromagnetic processes in the GEANT code,
so that each time the photon interacts with the matter (via absorption or scattering) the hidden photon is produced 
with probability given by the vacuum oscillation formula~\eqref{eq:osc} multiplied by the factor
{\it one half} as explained above. The
results are presented in Fig.~\ref{fig:dp_GEANT}, 
where we show the spectrum of photons produced by a 2.9~GW nuclear
reactor as well as the expected spectrum of dark photons calculated
without secondary photons (i.e. applying~\eqref{eq:dark}), using
the enhancement factor $f(E)$ which takes into account Compton-scattered
photons only and using full GEANT simulation. Here we take
$m_X=300$~eV and $\epsilon=0.01$ as an example. We observe that
the enhancement factor $f(E)$ perfectly describes the results of 
GEANT simulations apart from the very low energy region where there is 
a sharp spike at $E\approx0.5$~MeV which is attributed to
contributions from $e^+e^-$ annihilation process.

{\bf 5.} To summarize, we confirm the amplification of the dark photon production
via mixing with ordinary photon due to the secondary photons. However,
the overall enhancement is considerably smaller than
 that claimed in~\cite{Du:2023zlt}. The enhancement is noticeable only at low energies: the amplification factor grows from 1.2 to 2.6 with energy decrease from 2\,MeV to 0.2\,MeV. The enhancement due to photon
scattering processes is of the same size and shape for other masses of
dark photons. We expect that the sharp spike at $E\approx0.5$~MeV
observed in the expected spectrum of dark photons can serve as a
potential signature in future searches. 

We conclude with a comment about the spectrum of primary photons
for which in the present study (and following previous works \cite{Danilov:2018bks,Du:2023zlt}) we used the simple FRJ-1 research
reactor model~\cite{FRJ}. The effects, discussed in the main text, are exhibited only at low energies and numerically do not exceed 
a factor 2-3, which is below the accuracy of the
model~\cite{FRJ}. Even most recent calculations (see,
e.g.~\cite{Fujio:2023gzj}) of prompt fission $\gamma$-ray spectra
(which is the main part of total photon flux) have uncertainty up to a
factor of few. Moreover, the primary photon spectrum should be calculated on a case-by-case basis to ensure not only specific active zone media but also the specific changes in the latter during the working cycle of the reactor (see, e.g. Refs.~\cite{Bui:2016otf,MINER:2016igy,Mirzakhani:2025bqz}).
The enhancement factor $f(E)$ discussed in the present
study can be easily found for any reactor media.
However, its application for further tightening the experimental
bounds on new physics models obviously requires a better knowledge of the primary reactor 
photon spectrum.  

\vskip 0.5cm 

We thank Dmitry Kalashnikov and Mikhail Skorokhvatov   for helpful discussions. 
The work of SD and DG was supported by the the Russian Science
Foundation grant No.25-12-00309. 


\bibliography{refs} 

@article{Du:2023zlt,
    author = "Du, Mingxuan and Liu, Jia and Wang, Xiao-Ping and Wu, Tianhao",
    title = "{Amplifying nonresonant production of dark sector particles in scattering dominance regime}",
    eprint = "2309.00231",
    archivePrefix = "arXiv",
    primaryClass = "hep-ph",
    doi = "10.1103/PhysRevD.109.055041",
    journal = "Phys. Rev. D",
    volume = "109",
    number = "5",
    pages = "055041",
    year = "2024"
}

@article{Fabbrichesi:2020wbt,
    author = "Fabbrichesi, Marco and Gabrielli, Emidio and Lanfranchi, Gaia",
    title = "{The Dark Photon}",
    eprint = "2005.01515",
    archivePrefix = "arXiv",
    primaryClass = "hep-ph",
    doi = "10.1007/978-3-030-62519-1",
    month = "5",
    year = "2020"
}

@article{Danilov:2018bks,
    author = "Danilov, Mikhail and Demidov, Sergey and Gorbunov, Dmitry",
    title = "{Constraints on hidden photons produced in nuclear reactors}",
    eprint = "1804.10777",
    archivePrefix = "arXiv",
    primaryClass = "hep-ph",
    reportNumber = "INR-TH-2018-009",
    doi = "10.1103/PhysRevLett.122.041801",
    journal = "Phys. Rev. Lett.",
    volume = "122",
    number = "4",
    pages = "041801",
    year = "2019"
}

@article{Demidov:2018odn,
    author = "Demidov, S. and Gninenko, S. and Gorbunov, D.",
    title = "{Light hidden photon production in high energy collisions}",
    eprint = "1812.02719",
    archivePrefix = "arXiv",
    primaryClass = "hep-ph",
    reportNumber = "INR-TH-2018-029",
    doi = "10.1007/JHEP07(2019)162",
    journal = "JHEP",
    volume = "07",
    pages = "162",
    year = "2019"
}

@article{Seo:2020dtx,
    author = "Seo, S. H. and Kim, Y. D.",
    title = "{Dark Photon Search at Yemilab, Korea}",
    eprint = "2009.11155",
    archivePrefix = "arXiv",
    primaryClass = "hep-ph",
    doi = "10.1007/JHEP04(2021)135",
    journal = "JHEP",
    volume = "04",
    pages = "135",
    year = "2021"
}

@article{NEON:2024bpw,
    author = "Choi, J. J. and others",
    collaboration = "NEON",
    title = "{First Direct Search for Light Dark Matter Using the NEON Experiment at a Nuclear Reactor}",
    eprint = "2407.16194",
    archivePrefix = "arXiv",
    primaryClass = "hep-ex",
    doi = "10.1103/PhysRevLett.134.021802",
    journal = "Phys. Rev. Lett.",
    volume = "134",
    number = "2",
    pages = "021802",
    year = "2025"
}

@article{Caputo:2021eaa,
    author = "Caputo, Andrea and Millar, Alexander J. and O'Hare, Ciaran A. J. and Vitagliano, Edoardo",
    title = "{Dark photon limits: A handbook}",
    eprint = "2105.04565",
    archivePrefix = "arXiv",
    primaryClass = "hep-ph",
    reportNumber = "NORDITA-2021-036",
    doi = "10.1103/PhysRevD.104.095029",
    journal = "Phys. Rev. D",
    volume = "104",
    number = "9",
    pages = "095029",
    year = "2021"
}

@article{Holdom:1985ag,
    author = "Holdom, Bob",
    title = "{Two U(1)'s and Epsilon Charge Shifts}",
    reportNumber = "UTPT-85-30",
    doi = "10.1016/0370-2693(86)91377-8",
    journal = "Phys. Lett. B",
    volume = "166",
    pages = "196--198",
    year = "1986"
}

@article{Redondo:2015iea,
    author = "Redondo, Javier",
    title = "{Atlas of solar hidden photon emission}",
    eprint = "1501.07292",
    archivePrefix = "arXiv",
    primaryClass = "hep-ph",
    reportNumber = "MPP-2014-33",
    doi = "10.1088/1475-7516/2015/07/024",
    journal = "JCAP",
    volume = "07",
    pages = "024",
    year = "2015"
}

@article{Redondo:2008aa,
    author = "Redondo, Javier",
    title = "{Helioscope Bounds on Hidden Sector Photons}",
    eprint = "0801.1527",
    archivePrefix = "arXiv",
    primaryClass = "hep-ph",
    reportNumber = "DESY-07-211",
    doi = "10.1088/1475-7516/2008/07/008",
    journal = "JCAP",
    volume = "07",
    pages = "008",
    year = "2008"
}

@article{Redondo:2013lna,
    author = "Redondo, Javier and Raffelt, Georg",
    title = "{Solar constraints on hidden photons re-visited}",
    eprint = "1305.2920",
    archivePrefix = "arXiv",
    primaryClass = "hep-ph",
    reportNumber = "MPP-2013-110",
    doi = "10.1088/1475-7516/2013/08/034",
    journal = "JCAP",
    volume = "08",
    pages = "034",
    year = "2013"
}

@article{Braaten:1993jw,
    author = "Braaten, Eric and Segel, Daniel",
    title = "{Neutrino energy loss from the plasma process at all temperatures and densities}",
    eprint = "hep-ph/9302213",
    archivePrefix = "arXiv",
    reportNumber = "NUHEP-TH-93-1",
    doi = "10.1103/PhysRevD.48.1478",
    journal = "Phys. Rev. D",
    volume = "48",
    pages = "1478--1491",
    year = "1993"
}

@article{Park:2017prx,
    author = "Park, HyangKyu",
    title = "{Detecting Dark Photons with Reactor Neutrino Experiments}",
    eprint = "1705.02470",
    archivePrefix = "arXiv",
    primaryClass = "hep-ph",
    doi = "10.1103/PhysRevLett.119.081801",
    journal = "Phys. Rev. Lett.",
    volume = "119",
    number = "8",
    pages = "081801",
    year = "2017"
}

@article{FRJ,
      author         = "Bechteler, H and others",
      title          = "{ The spectrum of $\gamma$ radiation emitted in the FRJ-1 (Merlin) reactor core and moderator region}",
      journal        = "Spezielle Berichte der Kernforschungsanlage Juelich",
      volume         = "255",
      year           = "1984",
      number         = "62",
      pages          = "62"
}

@article{ParticleDataGroup:2024cfk,
    author = "Navas, S. and others",
    collaboration = "Particle Data Group",
    title = "{Review of particle physics}",
    doi = "10.1103/PhysRevD.110.030001",
    journal = "Phys. Rev. D",
    volume = "110",
    number = "3",
    pages = "030001",
    year = "2024"
}

@misc{148746,
    author = "Martin Berger and J Hubbell and Stephen Seltzer and J Coursey and D Zucker",
    title = "XCOM: Photon Cross Section Database (version 1.2)",
    publisher = "http://physics.nist.gov/xcom",
    year = "1999"
}

@book{LL4,
    author = "V.B.Berestetskii and E.M.Lifshitz and L.P.Pitaevskii",
    title = "Relativistic Quantum Theory. Vol. 4 (1st ed.)",
    publisher = "Pergamon Press.",
    isbn = "978-0-08-017175-3",
    year = "1971"
}

@article{Sigl:1993ctk,
    author = "Sigl, G. and Raffelt, G.",
    title = "{General kinetic description of relativistic mixed neutrinos}",
    reportNumber = "MPI-PH-92-112",
    doi = "10.1016/0550-3213(93)90175-O",
    journal = "Nucl. Phys. B",
    volume = "406",
    pages = "423--451",
    year = "1993"
}

@article{Stodolsky:1986dx,
    author = "Stodolsky, Leo",
    title = "{On the Treatment of Neutrino Oscillations in a Thermal Environment}",
    reportNumber = "MPI-PAE/PTh 77/86",
    doi = "10.1103/PhysRevD.36.2273",
    journal = "Phys. Rev. D",
    volume = "36",
    pages = "2273",
    year = "1987"
}

@article{Thomson:1991xq,
    author = "Thomson, Mark J.",
    title = "{The Damping of quantum coherence by elastic and inelastic processes}",
    reportNumber = "M-C-TH-91-13, MC-TH-91-13",
    doi = "10.1103/PhysRevA.45.2243",
    journal = "Phys. Rev. A",
    volume = "45",
    pages = "2243--2249",
    year = "1992"
}

@article{Foot:1996qc,
    author = "Foot, Robert and Volkas, R. R.",
    title = "{Studies of neutrino asymmetries generated by ordinary sterile neutrino oscillations in the early universe and implications for big bang nucleosynthesis bounds}",
    eprint = "hep-ph/9610229",
    archivePrefix = "arXiv",
    reportNumber = "UM-P-96-81, RCHEP-96-09",
    doi = "10.1103/PhysRevD.55.5147",
    journal = "Phys. Rev. D",
    volume = "55",
    pages = "5147--5176",
    year = "1997"
}

@article{Fujio:2023gzj,
    author = {Fujio, Kazuki and Al-Adili, Ali and Nordstr{\"o}m, Fredrik and Lema{\^\i}tre, Jean-Fran{\c{c}}ois and Okumura, Shin and Chiba, Satoshi and Koning, Arjan},
    title = "{TALYS calculations of prompt fission observables and independent fission product yields for the neutron-induced fission of $^{235}$U}",
    doi = "10.1140/epja/s10050-023-01095-4",
    journal = "Eur. Phys. J. A",
    volume = "59",
    number = "8",
    pages = "178",
    year = "2023"
}

@article{Bui:2016otf,
    author = "Bui, V. M. and others",
    title = "{Antineutrino emission and gamma background characteristics from a thermal research reactor}",
    eprint = "1602.07522",
    archivePrefix = "arXiv",
    primaryClass = "nucl-ex",
    month = "2",
    year = "2016"
}

@article{MINER:2016igy,
    author = "Agnolet, G. and others",
    collaboration = "MINER",
    title = "{Background Studies for the MINER Coherent Neutrino Scattering Reactor Experiment}",
    eprint = "1609.02066",
    archivePrefix = "arXiv",
    primaryClass = "physics.ins-det",
    doi = "10.1016/j.nima.2017.02.024",
    journal = "Nucl. Instrum. Meth. A",
    volume = "853",
    pages = "53--60",
    year = "2017"
}

@article{Mirzakhani:2025bqz,
    author = "Mirzakhani, M. and others",
    title = "{MINER reactor based search for axionlike particles using sapphire (Al2O3) detectors}",
    eprint = "2504.20960",
    archivePrefix = "arXiv",
    primaryClass = "hep-ex",
    doi = "10.1103/tbwx-b5lx",
    journal = "Phys. Rev. D",
    volume = "112",
    number = "3",
    pages = "032013",
    year = "2025"
}


\end{document}